\documentclass{article}
\usepackage{spconf,amsmath,graphicx,hyperref, enumitem, booktabs, amsmath, amssymb, cite, multirow, multicol, caption}
\usepackage[dvipsnames]{xcolor}

\usepackage{booktabs}
\usepackage{multirow}
\usepackage{siunitx} 
\usepackage[table]{xcolor} 
\usepackage{threeparttable} 

\title{
Sounding Highlights: Dual-Pathway Audio Encoders\\for Audio-Visual Video Highlight Detection}

\name{Seohyun Joo$^{1\textasteriskcentered}$ \thanks{$\textasteriskcentered$ This work was done during an internship at Music and Audio Research Group, Seoul National University.} \qquad Yoori Oh$^{2}$ }
  
\address{
    $^{1}$ School of Electrical Engineering and Computer Science, GIST \\
    $^{2}$ Music and Audio Research Group, Seoul National University \\
    seohyunj@gm.gist.ac.kr, yoori0203@snu.ac.kr
}

\begin{document}
\ninept
\maketitle
\begin{abstract}
\vspace{-2pt}

Audio-visual video highlight detection aims to automatically identify the most salient moments in videos by leveraging both visual and auditory cues. However, existing models often underutilize the audio modality, focusing on high-level semantic features while failing to fully leverage the rich, dynamic characteristics of sound. To address this limitation, we propose a novel framework, Dual-Pathway Audio Encoders for Video Highlight Detection (DAViHD). The dual-pathway audio encoder is composed of a semantic pathway for content understanding and a dynamic pathway that captures spectro-temporal dynamics. The semantic pathway extracts high-level information by identifying the content within the audio, such as speech, music, or specific sound events. The dynamic pathway employs a frequency-adaptive mechanism as time evolves to jointly model these dynamics, enabling it to identify transient acoustic events via salient spectral bands and rapid energy changes. We integrate the novel audio encoder into a full audio-visual framework and achieve new state-of-the-art performance on the large-scale Mr.HiSum benchmark. Our results demonstrate that a sophisticated, dual-faceted audio representation is key to advancing the field of highlight detection.

\end{abstract}
\begin{keywords}
Video Highlight Detection, Audio-Visual Learning,  Multimodal Fusion, Spectro-Temporal Dynamics
\vspace{-4pt}
\end{keywords}

\section{Introduction}
\label{sec:intro}
\vspace{-5pt}


With the rapid growth of digital video, automated video understanding has become an essential area of research. A key task within this field is video highlight detection, which focuses on identifying the most interesting and engaging moments from a video. This capability is fundamental for applications such as effective content summarization, retrieval, and recommendation.

\begin{figure}[!t]
    \centerline{\includegraphics[width=\columnwidth]{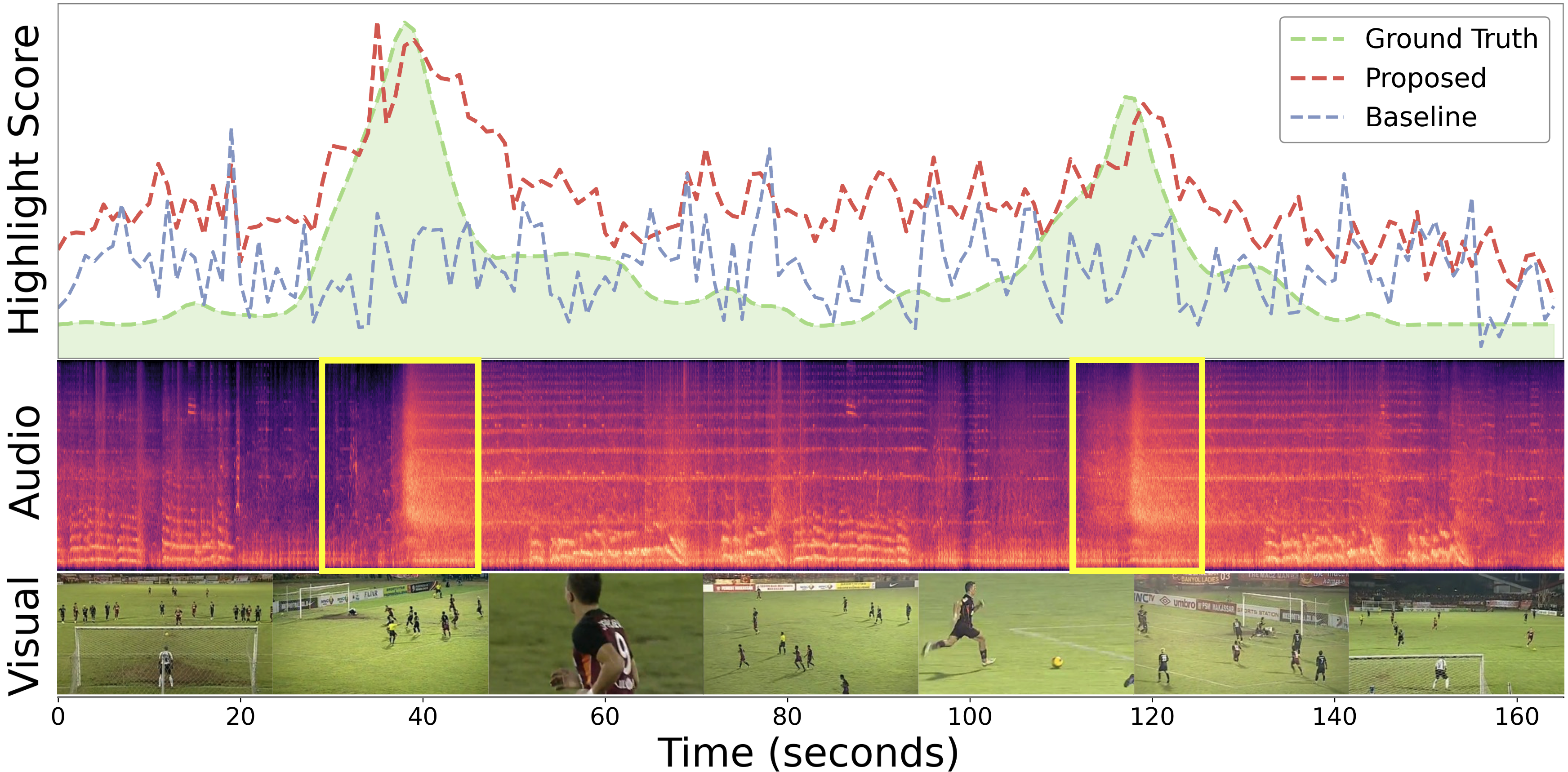}}
    \vspace{-8pt}
    \caption{Comparison of baseline and proposed framework. {\color{blue} The baseline model}\cite{jointav} produces uniform scores by relying on global audio-visual features, failing to match the {\color{Green}ground-truth}. {\color{red} Our proposed} framework, however, accurately captures the {\color{Green}ground-truth} dynamics by utilizing abrupt auditory changes, highlighted by the yellow boxes in the audio, as key features.}
    \label{fig:motivation}
    \vspace{-15pt}
\end{figure}

Initial research in video highlight detection primarily centered on visual-only approaches like PGL-SUM~\cite{pglsum}. Subsequent work began to integrate audio, with methods like Joint-VA~\cite{jointav} employing bimodal attention to model cross-modal interactions. More recently, the field has shifted towards large-scale, Transformer-based architectures. While models like VATT~\cite{vatt}, MBT~\cite{mbt}, and UMT~\cite{umt} introduced sophisticated multimodal fusion mechanisms, they often treat audio simplistically, relying on generic high-level semantic features from pre-trained backbones like PANNs~\cite{panns} while overlooking its rich spectro-temporal dynamics. This represents a significant gap, as the acoustic properties of sound itself are critical for identifying salient moments. The significance of these dynamics is particularly apparent in video highlight detection in Figure~\ref{fig:motivation}. Baseline approaches (blue) are insensitive to the transient acoustic events that signal key moments due to their reliance on general audio-visual information. Our framework (red), however, leverages the abrupt auditory changes visualized in the spectrogram's yellow boxes as crucial clues. By modeling these specific dynamics, it can predict highlight scores that closely mirror the green ground-truth pattern.

To address this, we draw inspiration from advances in audio representation learning. Early convolution-based methods~\cite{avslowfast, sfauditstream} explored temporal properties but were limited by the spatial equivalence assumption of 2D convolutions. A key breakthrough was Frequency-Dynamic Convolution (FDC)~\cite{nam2022frequency}, which was proposed to address the physical inconsistencies of applying standard CNNs to spectrograms. Building on subsequent refinements of FDC that aim to capture key spectro-temporal information for transient acoustic events~\cite{nam2024diversifying, nam2025temporal}, we propose a framework that utilizes two critical facets of sound: the semantic content of sound (\textit{i.e., what} is heard) and its spectro-temporal dynamics (\textit{i.e., how} the sound evolves).

To this end, we introduce a novel framework, Dual-Pathway Audio Encoders for Video Highlight Detection (DAViHD). Our approach explicitly models both the high-level semantic content and the low-level spectro-temporal dynamics of the audio signal. The main contributions of our work are as follows:
\begin{itemize}[leftmargin=10pt]
    \vspace{-4pt}
    \item We propose the Dual-Pathway Audio Encoder, an architecture that disentangles the audio signal into two distinct streams to model both its semantic and dynamic properties.
    \vspace{-4pt}
    \item By integrating our audio encoder into a unified audio-visual framework, we achieve state-of-the-art (SOTA) performance on the large-scale Mr.HiSum~\cite{mrhisum} benchmark.
    \vspace{-4pt}
\end{itemize}
The demo is available at \url{https://seohyj.github.io/soundhd.github.io/}.

\begin{figure*}[!t]
    \centering
    \includegraphics[width=1\textwidth]{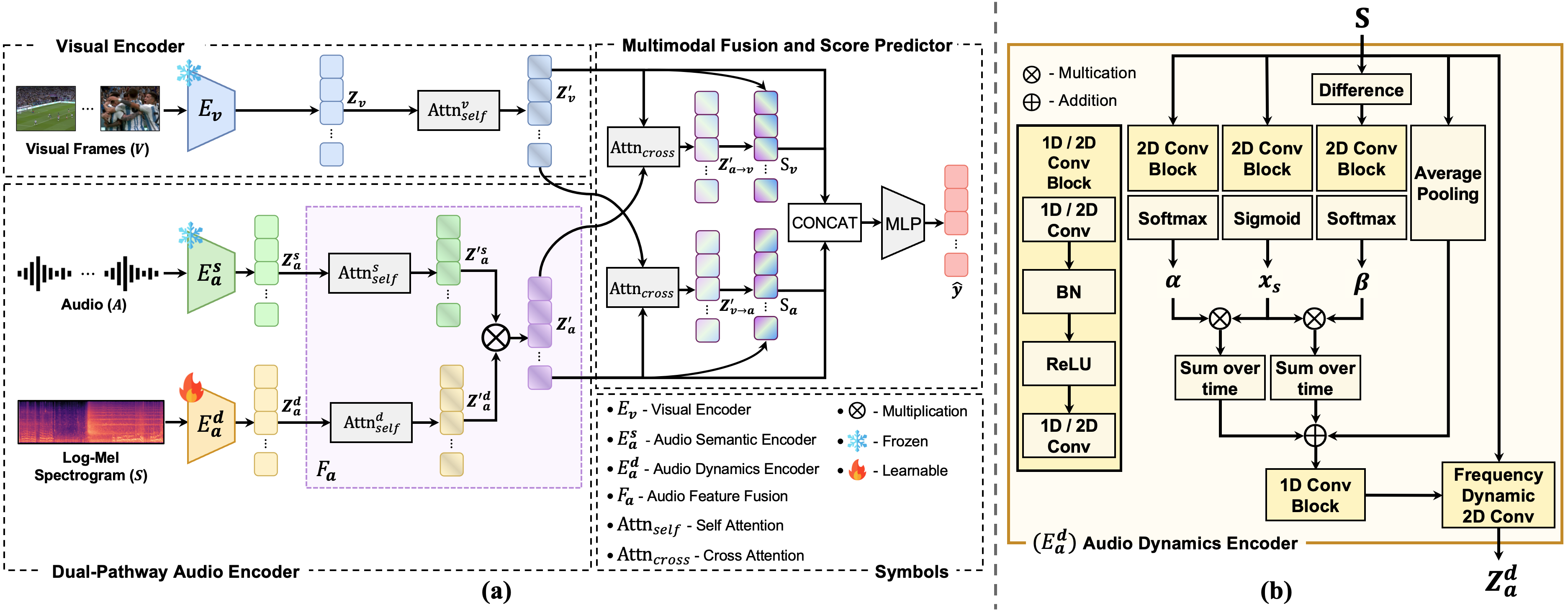} 
    \vspace{-16pt}
    \caption{\textbf{(a)} An overview of the DAViHD framework. The model comprises a Visual Encoder ($E_v$) and a Dual-Pathway Audio Encoder ($E_a^s, E_a^d$). Features from both audio encoders are fused via the Audio Feature Fusion module ($F_a$), and the fused audio feature, $\mathbf{Z}'_a$, is passed to a cross-attention module and an MLP to predict the final score $\hat{y}$. \textbf{(b)} Detailed architecture of the Audio Dynamics Encoder ($E_a^d$). It uses a multi-branch architecture to fuse two attention maps ($\alpha, \beta$), modulated by a saliency gate ($x_s$), with a global average-pooled feature. This information is then used to dynamically control a Frequency-Dynamic convolutional layer.}
    \label{fig:overall_arch}
    \vspace{-10pt}
\end{figure*}


\section{Methods}
\label{sec:format}
\vspace{-4pt}

Figure~\ref{fig:overall_arch} illustrates the architecture of our proposed DAViHD framework. The framework  processes visual and audio streams using a visual encoder and our novel dual-pathway audio encoder, respectively. The extracted multi-modal features are subsequently fused and fed into a score predictor to yield the final highlight scores.

\vspace{-6pt}
\subsection{Visual Encoder}
\label{ssec:subhead}
\vspace{-3pt}

The visual stream takes a sequence of video frames $\mathbf{V} \in \mathbb{R}^{T_f \times H \times W \times C}$ as input, where $T_f$, $H$, $W$, and $C$ denote the number of frames, height, width, and channels, respectively. Initially, a pre-trained convolutional neural network, $E_v$, is employed as the visual backbone~\cite{resnet,inceptionv3} to extract a sequence of frame-level features, $\mathbf{Z}_v = E_v(\mathbf{V})$. The resulting feature sequence $\mathbf{Z}_v \in \mathbb{R}^{T_f \times D_{v}}$, where $D_{v}$ is the visual embedding dimension, is then processed by a multi-head self-attention mechanism \cite{transformer} to capture long-range contextual dependencies, yielding the final visual representation $\mathbf{Z'}_v \in \mathbb{R}^{T_f \times D_v}$.

\vspace{-6pt}
\subsection{Dual-Pathway Audio Encoder}
\label{ssec:dual_pathway}
\vspace{-3pt}

As illustrated in Figure~\ref{fig:overall_arch}(a), the dual-pathway audio encoder extracts a comprehensive audio representation by processing the audio stream through two parallel pathways. 
The first pathway, an audio semantic encoder, processes the raw audio waveform in 1-second segments to model high-level semantic properties. 
Concurrently, the second pathway, an audio dynamics encoder, takes a log-Mel spectrogram as input to capture low-level dynamic properties. 
The outputs from these pathways, audio semantic embeddings $\mathbf{Z}_a^s \in \mathbb{R}^{T_f \times D_s}$ and audio dynamics features $\mathbf{Z}_a^d \in \mathbb{R}^{T_f \times D_d}$, are then fused by an audio feature fusion module. 
This creates a unified feature sequence $\mathbf{Z}_a \in \mathbb{R}^{T_f \times D_a}$, where $D_s, D_d,$ and $D_a$ are the respective dimensions of the semantic, dynamic, and fused audio embeddings.


\vspace{-6pt}
\subsubsection{Audio Semantic Encoder}
\label{ssec:semantic_encoder}
\vspace{-3pt}

The audio semantic encoder, $E_a^s$, is designed to interpret the high-level meaning of the audio signal, essentially, \textit{what} is being heard. To achieve this, we employ the pre-trained PANN model~\cite{panns}, which takes the raw audio waveform $\mathbf{A} \in \mathbb{R}^{L}$ as input, where $L$ denotes the number of samples. The model first segments the waveform into a sequence of non-overlapping 1-second chunks. Each chunk is then processed independently by the encoder to extract a high-dimensional semantic embedding. Finally, these segment-level embeddings are concatenated chronologically to form the final sequence of semantic embeddings $\mathbf{Z}_a^s \in \mathbb{R}^{T_f \times D_s}$. We denote this entire procedure as $\mathbf{Z}_a^s = E_a^s(\mathbf{A})$.

\vspace{-6pt}
\subsubsection{Audio Dynamics Encoder}
\label{ssec:dynamics_encoder}
\vspace{-3pt}


A standard convolutional block is utilized throughout the model, termed the 1D/2D Conv block. It is defined as a sequence of a Convolutional layer (1D or 2D), Batch Normalization, a ReLU activation, and a final, dimension-matching Convolutional layer.
The audio dynamics encoder, $E_a^d$, is designed to capture low-level, time-varying patterns from a log-Mel spectrogram, $\mathbf{S} \in \mathbb{R}^{F \times T}$, where $F$ is the number of frequency bins and $T$ is the total number of time frames indexed by $t$. The encoder generates a final audio dynamics feature, $\mathbf{Z}_a^d = E_a^d(\mathbf{S})$.
\vspace{1pt}
As shown in Figure \ref{fig:overall_arch}(b), it processes the input through parallel branches.  
Two branches take the log-Mel spectrogram $\mathbf{S}$ as input and use 2D Conv blocks to generate a temporal attention map  $\boldsymbol{\alpha}$ via a softmax function and a saliency gate  $\boldsymbol{x_s}$ via a sigmoid function. A third branch captures temporal changes by processing the frame-to-frame difference, $\Delta\mathbf{S} = |\mathbf{S}_t - \mathbf{S}_{t-1}|$, which is calculated by a Difference module. $\Delta\mathbf{S}$ is then passed through the branch's own 2D Conv block and a softmax function to produce a velocity attention map, $\boldsymbol{\beta}$.
These maps are then used to compute time-aware features ($\mathbf{f}_{\text{TA}}$) and velocity-aware features ($\mathbf{f}_{\text{VA}}$):
\begingroup
\setlength{\belowdisplayskip}{0pt}
\setlength{\belowdisplayshortskip}{0pt}
\begin{align}
    \mathbf{f}_{\text{TA}} &= \sum_{t} (\boldsymbol{\alpha} \otimes \mathbf{x_s}), \quad\quad
    \mathbf{f}_{\text{VA}} = \sum_{t} (\boldsymbol{\beta} \otimes \mathbf{x_s}), 
\end{align}
\endgroup
where $\otimes$ denotes element-wise multiplication. These features are then summed with a global context vector from a fourth average pooling branch to form a final combined vector, $\mathbf{f}_{\text{combined}}$.
This vector controls a Frequency-Dynamic 2D Conv layer~\cite{nam2022frequency}. 
The combined vector $\mathbf{f}_{\text{combined}}$ is passed through a 1D Conv Block to generate frequency-specific weights $\gamma_k(f)$, where $f$ denotes a specific frequency bin, for a set of $K$ learnable basis kernels $\{\mathbf{W}_k\}_{k=1}^K$. These weights modulate the basis kernels, and the resulting dynamic filter is applied to the original input spectrogram $\mathbf{S}$. The final audio dynamics feature $\mathbf{Z}_a^d$ is computed as the following Eq.(~\ref{eq:tfd_conv}), where $*$ denotes the 2D convolution. 
\vspace{-3pt}
\begin{equation}
    \mathbf{Z}_a^d = \sum_{k=1}^{K} \gamma_k \otimes (\mathbf{W}_k * \mathbf{S}).
    \label{eq:tfd_conv}
\end{equation}
This mechanism, unlike standard 2D convolutions that treat the time and frequency axes as spatially equivalent, allows the layer to form an adaptive filter by dynamically combining the outputs of the fixed basis kernels. 
The feature map output from the Frequency Dynamic 2D Conv layer is transformed into the final sequence representation. The channel and frequency dimensions of the feature map are first flattened, then temporally aligned with the other modalities via adaptive average pooling to match the temporal length of $T_f$. Finally, a linear projection layer maps the sequence to the target embedding dimension, yielding the final audio dynamics representation $\mathbf{Z}_a^d$.

\vspace{-6pt}
\subsubsection{Audio Feature Fusion}
\label{ssec:audio_fusion}
\vspace{-3pt}

The audio feature fusion module $F_a$  in Figure~\ref{fig:overall_arch}(a) integrates the  representations from the semantic and dynamics pathways. To contextualize each stream prior to fusion, the initial semantic feature, $\mathbf{Z}_{a}^{s}$, and dynamics feature, $\textbf{Z}_{a}^{d}$, are each passed through self-attention \cite{transformer}. This approach of applying self-attention before fusion, which we term Early-SA, allows the model to independently capture the temporal dependencies in each stream.
As will be demonstrated in our ablation study (Section~\ref{ssec:ablation_study}), this prior contextualization is crucial for effective fusion. This operation yields refined, context-aware representations $\mathbf{Z}_{a}^{\prime s}=\text{Attn}_{self}^s(\mathbf{Z}_{a}^{s})$ and $\mathbf{Z}_{a}^{\prime d}=\text{Attn}_{self}^d(\mathbf{Z}_{a}^{d})$, where $\text{Attn}_{self}^s$ and $\text{Attn}_{self}^d$ denote two independent self-attention blocks with their own learnable parameters. These contextualized features are then fused using element-wise multiplication. This fusion strategy acts as a gating mechanism, allowing the low-level dynamics features ($\mathbf{Z}_{a}^{\prime d}$) to modulate the high-level semantic features ($\mathbf{Z}_{a}^{\prime s}$) and amplify the salient information. The final unified audio representation, $\mathbf{Z}_{a}^{\prime} \in \mathbb{R}^{T_f \times D_a}$, is computed as:
\begin{equation}
    \mathbf{Z}_{a}^{\prime} = \mathbf{Z}_{a}^{\prime s} \otimes \mathbf{Z}_{a}^{\prime d},
    \label{eq:audio_fusion}
\end{equation}
where $\otimes$ denotes element-wise multiplication. This representation is then passed directly to the multimodal fusion stage.


\vspace{-7pt}
\subsection{Multimodal Fusion and Score Prediction}
\label{ssec:fusion}
\vspace{-4pt}

The final stage of our framework fuses the context-aware visual and audio representation, $\mathbf{Z}_v^{\prime}$, and $\mathbf{Z}_a^{\prime}$. We employ a bidirectional cross-modal attention $\text{Attn}_{cross}$, following prior works \cite{vatt, jointav, mbt, umt, eclipse}, to allow each modality to attend to and incorporate features from the other. First, we compute an audio-contextualized visual representation, $\mathbf{Z}'_{a \to v}$, where the visual representations act as queries ($\mathbf{Q}_v$) and the audio representations provide the keys ($\mathbf{K}_a$) and values ($\mathbf{V}_a$):
\vspace{-2pt}
\begin{equation}
    \mathbf{Z}'_{a \to v} = \text{softmax}\left(\frac{\mathbf{Q}_v \mathbf{K}_a^T}{\sqrt{d_k}}\right)\mathbf{V}_a.
    \label{eq:cross_attention}
\end{equation}
Symmetrically, we compute a visual-contextualized audio representation, $\mathbf{Z}'_{v \to a}$, by swapping the roles of the audio and visual streams. To preserve original modality-specific information during this fusion, we then apply a residual connection \cite{resnet} to each stream, resulting in the enhanced representations $\mathbf{S}_v=\mathbf{Z}_v^{\prime} + \mathbf{Z}_{a \to v}^{\prime}$ and $ \mathbf{S}_a=\mathbf{Z}_a^{\prime} + \mathbf{Z}_{v \to a}^{\prime}$.

For the final prediction, the original self-attended features ($\mathbf{Z}'_v$, $\mathbf{Z}'_a$) and the enhanced cross-attended features ($\mathbf{S}_v$, $\mathbf{S}_a$) are all concatenated together. The resulting concatenated vector is processed by a 3-layer MLP which regresses the final frame-level highlight score, normalized between 0 and 1. The model is optimized by minimizing the Mean Squared Error (MSE) between the predicted scores $\hat{y}$ and the ground-truth scores $y$:
    \vspace{-5pt}
    \begin{equation}
        \mathcal{L}_{\text{MSE}} = \frac{1}{T} \sum_{t=1}^{T} (y_t - \hat{y}_t)^2.
    \end{equation}


\vspace{-12pt}
\section{Experiments}
\label{sec:pagestyle}

\vspace{-5pt}
\subsection{Experimental Setups}
\label{ssec:subhead}

\vspace{-3pt}
\subsubsection{Datasets}
\label{ssec:subhead}
\vspace{-3pt}

The performance is evaluated on the TVSum \cite{tvsum} and Mr.HiSum \cite{mrhisum} video highlight detection benchmarks. TVSum consists of 50 web videos from 10 diverse categories. Following standard evaluation protocols \cite{pglsum, lee2025video}, we use 5-fold cross-validation and report the average performance. Mr.HiSum is a large-scale dataset comprising 31,892 YouTube videos with an average length of 201.9 seconds. Its highlight scores are derived from YouTube's `Most replayed' statistics, providing a robust, user-driven measure. After filtering for unavailable videos, the experiments are conducted on 30,656 videos.

\vspace{-7pt}
\subsubsection{Evaluation Metrics}
\label{ssec:subhead}
\vspace{-3pt}

We use a set of metrics to assess different aspects of its performance. To measure temporal localization accuracy, we use the F1-score. Following standard evaluation protocols, we convert the continuous prediction scores into a binary summary by selecting the top 50\% of segments and measure its temporal intersection with the ground-truth summary. For ranking quality, we report mean Average Precision (mAP). We compute mAP at two thresholds, $\text{mAP}_{\rho=15\%}$ and $\text{mAP}_{\rho=50\%}$, which evaluate whether ground-truth highlight segments are ranked within the top 15\% and 50\% of all predictions.  Finally, to evaluate the alignment with human perception, we use Spearman’s $\rho$ and Kendall’s $\tau$ rank correlation coefficients. These metrics directly measure the monotonic agreement between the predicted score sequence and the ground-truth annotations.

\begin{table*}[!h]
\centering
\caption{
Main results and comparison on the Mr.HiSum and TVSum datasets. All metrics are reported as mean $\pm$ std. over 5 runs. Higher values indicate better performance, and the best results for each metric are highlighted in bold.}
\vspace{-8pt}
\label{tab:main_results}
\resizebox{\textwidth}{!}{
    \begin{tabular}{l ccccc ccccc}
    \toprule
    \multirow{2}{*}{Model}  
    & \multicolumn{5}{c}{Mr.HiSum} 
    & \multicolumn{5}{c}{TVSum} \\
    \cmidrule(lr){2-6} \cmidrule(lr){7-11}
    & F1 $\uparrow$ & mAP$_{\rho=50\%}$ $\uparrow$ & mAP$_{\rho=15\%}$ $\uparrow$ & $\rho$ $\uparrow$ & $\tau$ $\uparrow$ 
    & F1 $\uparrow$ & mAP$_{\rho=50\%}$ $\uparrow$ & mAP$_{\rho=15\%}$ $\uparrow$ & $\rho$ $\uparrow$ & $\tau$ $\uparrow$ \\
    \midrule
    PGL-SUM$^{\dagger}$\cite{pglsum}       
    & 53.34$\pm$0.10 & 59.73$\pm$0.17 & 25.71$\pm$0.30 & 0.104$\pm$0.003 & 0.070$\pm$0.002 
    & 52.93$\pm$1.75 & 56.68$\pm$2.33 & 23.18$\pm$1.96 & 0.056$\pm$0.040 & 0.038$\pm$0.027 \\
    CSTA$^{\dagger}$\cite{csta}       
    & 54.32$\pm$0.17 & 61.12$\pm$0.39 & 28.35$\pm$0.48 & 0.138$\pm$0.005 & 0.095$\pm$0.004 
    & 57.32$\pm$1.99 & 62.36$\pm$2.81 & 27.52$\pm$5.08 & \textbf{0.205}$\pm$\textbf{0.056} & \textbf{0.141}$\pm$\textbf{0.041} \\
    \midrule
    Joint-VA $^{\ddagger}$\cite{jointav} 
    & 54.71$\pm$0.04 & 61.82$\pm$0.11 & 29.09$\pm$0.22 & 0.152$\pm$0.001 & 0.104$\pm$0.001 
    & 55.03$\pm$2.20 & 60.94$\pm$3.19 & 26.66$\pm$3.40 & 0.142$\pm$0.046 & 0.097$\pm$0.031 \\
    UMT$^{\ddagger}$\cite{umt}              
    & 58.18$\pm$0.29 & 65.81$\pm$0.31 & 33.79$\pm$0.35 & 0.239$\pm$0.006 & 0.174$\pm$0.004 
    & 57.54$\pm$0.87 & 61.49$\pm$2.91 & 25.24$\pm$5.05 & 0.175$\pm$0.022 & 0.121$\pm$0.015 \\
    \midrule
    \textbf{DAViHD (Ours)}$^{\ddagger}$ 
    & \textbf{59.73}$\pm$\textbf{0.41} & \textbf{67.27}$\pm$\textbf{0.52} & \textbf{36.55}$\pm$\textbf{0.51} & \textbf{0.299}$\pm$\textbf{0.012} & \textbf{0.213}$\pm$\textbf{0.009} 
    & \textbf{57.67}$\pm$\textbf{1.27} & \textbf{63.52}$\pm$\textbf{2.58} & \textbf{28.94}$\pm$\textbf{3.11} & 0.200$\pm$0.032 & 0.138$\pm$0.022 \\
    \bottomrule
    \end{tabular}
}
    \caption*{\raggedright\footnotesize{$^{\dagger}$ video only, $^{\ddagger}$ video and audio}}
    \vspace{-20pt}
\end{table*}

\vspace{-9pt}
\subsubsection{Implementation Details}
\label{ssec:implementation_details}
\vspace{-4pt}

All input videos are processed into 1-second segments at a rate of 1~fps.
For the visual stream, the feature dimension $D_v$ is adapted for each benchmark to align with standard practices. On the TVSum dataset, we follow prior work \cite{tvsum_ref1, jointav} and extract 512-dimensional visual features ($D_v = 512$) for each segment using a 3D CNN~\cite{3dcnn} with a ResNet-34~\cite{resnet} backbone pre-trained on Kinetics-400~\cite{kin400}. For the Mr.HiSum dataset~\cite{mrhisum}, we adhere to its standard protocol, extracting 1024-dimensional visual features ($D_v = 1024$) from an Inception-v3~\cite{inceptionv3} model pre-trained on ImageNet~\cite{imagenet}, followed by PCA.
For the audio stream, our dual-pathway audio encoder processes two parallel inputs. The audio semantic encoder utilizes PANNs~\cite{panns} pre-trained on AudioSet~\cite{audioset} to extract 2048-dimensional semantic embeddings ($D_s = 2048$). The Dynamic Pathway takes a log-Mel spectrogram (16~kHz, 2048 FFT, 256 hop, 128 mel frequency bins) as input. This spectrogram is fed into the Audio Dynamics Encoder(~\ref{ssec:dynamics_encoder}), which internally downsamples the feature representation using $K=4$ learnable basis kernels, to a final resolution of 1~fps to align with the semantic features, yielding a 2048-dimensional dynamic feature sequence ($D_d = 2048$). Finally, the semantic and dynamic features are fused, resulting in a unified audio representation with a final dimension of $D_a = 2048$.
Our model is trained using the Adam optimizer~\cite{adam}. For the Mr.HiSum dataset, we train for 200 epochs with a learning rate of $1 \times 10^{-5}$ and a batch size of 16. For TVSum dataset, we train for 400 epochs with a learning rate of $5 \times 10^{-6}$ and a batch size of 8. For both datasets, we use a weight decay of $1 \times 10^{-4}$ and apply gradient clipping with a maximum norm of 0.5.

\begin{table}[!t]
\centering
\vspace{-6pt}
\caption{Ablation study results for each modality's contribution. $V$ denotes the visual modality, while $A_s$ and $A_d$ represent the audio semantic and dynamics pathways, respectively.}
\vspace{-6pt}
\label{tab:ablation_modality}
\resizebox{\columnwidth}{!}{%
    \begin{tabular}{ccc ccccc}
    \toprule
    V & $A_s$ & $A_d$ & F1 $\uparrow$ & mAP$_{\rho=50\%}$ $\uparrow$ & mAP$_{\rho=15\%}$ $\uparrow$ & $\rho$ $\uparrow$ & $\tau$ $\uparrow$ \\
    \midrule
    \checkmark &            &            & 52.98 & 58.93 & 25.31 & 0.101 & 0.069 \\
               & \checkmark &            & 53.25 & 60.11 & 28.21 & 0.109 & 0.075 \\
               &            & \checkmark & 57.53 & 63.88 & 33.15 & 0.244 & 0.175 \\
    \midrule
    \checkmark & \checkmark &            & 54.79 & 61.95 & 28.94 & 0.153 & 0.105 \\
    \checkmark &            & \checkmark & 58.25 & 65.84 & 35.51 & 0.269 & 0.191 \\
               & \checkmark & \checkmark & 59.09 & 66.12 & 35.62 & 0.282 & 0.203 \\
    \midrule
    \checkmark & \checkmark & \checkmark & \textbf{60.17} & \textbf{68.01} & \textbf{36.96} & \textbf{0.312} & \textbf{0.224} \\
    \bottomrule
    \end{tabular}%
}
\vspace{-8pt}
\end{table}


\vspace{-10pt}
\subsection{Main Results}
\label{ssec:main_results}
\vspace{-5pt}

Table~\ref{tab:main_results} compares the performance of the proposed DAViHD model against several highlight detection methods, including PGL-SUM~\cite{pglsum}, Joint-VA~\cite{jointav}, UMT~\cite{umt} and CSTA~\cite{csta}. The results indicate that our model outperforms the baseline models in most evaluation metrics, achieving SOTA performance especially on the large-scale Mr.HiSum dataset.
To demonstrate the benefits of the audio modality, the DAViHD model is compared against a strong vision-only model (PGL-SUM and CSTA). Furthermore, it is benchmarked against two recent audio-visual models (Joint-VA and UMT). Joint-VA is considered a primary baseline due to its structural similarity to our approach. For UMT, its text-search functionality was disabled during evaluation to ensure a fair comparison.
To ensure the reliability of the results, each experiment was conducted five times with different random initializations, and the reported scores represent the average of these runs.
Notably, the proposed model outperforms even the powerful audio-visual baselines. This result highlights the effectiveness of the dual-pathway audio encoder in extracting and utilizing rich, informative features from the audio stream to identify salient moments in videos. A detailed analysis of each model component is presented in the following section.

\begin{table}[!t]
\centering
\caption{Ablation study on different audio fusion strategies. We analyze the impact of the self-attention (SA) layer placement (Early vs. Late) and the feature combination method (Concat vs. Multiply).}
\vspace{-6pt}
\label{tab:ablation_fusion}
\resizebox{\columnwidth}{!}{%
\begin{tabular}{l l ccccc}
\toprule
SA Placement & Combination  & F1 $\uparrow$ & mAP$_{\rho=50\%}$ $\uparrow$ & mAP$_{\rho=15\%}$ $\uparrow$ & $\rho$ $\uparrow$ & $\tau$ $\uparrow$ \\
\midrule
Late & Concat   & 58.71 & 66.24 & 35.61 & 0.280 & 0.198 \\
Late & Multiply  & 58.40 & 66.01 & 35.93 & 0.276 & 0.195 \\
\midrule
Early & Concat   & 59.42 & 67.36 & 36.21 & 0.294 & 0.208 \\
\textbf{Early} & \textbf{Multiply}  & \textbf{60.17} & \textbf{68.01} & \textbf{36.96} & \textbf{0.312} & \textbf{0.224} \\
\bottomrule
\end{tabular}%
}
\vspace{-15pt}
\end{table}

\vspace{-10pt}
\subsection{Ablation Studies}
\label{ssec:ablation_study}
\vspace{-5pt}

We conduct a series of ablation studies to analyze the contribution of each component in our proposed framework. All ablation experiments are performed on the large-scale, Mr.HiSum dataset.

\vspace{0pt}
\textbf{Contribution of each Modality} \; We evaluate the effectiveness of each input modality in an ablation study.
As shown in Table~\ref{tab:ablation_modality}, our ablation study reveals several key insights into the contribution of each modality.
First, when considering single streams, the audio dynamics pathway ($A_d$) alone significantly outperforms configurations using only the visual ($V$) or audio semantic ($A_s$) streams, demonstrating that spectro-temporal dynamics are highly effective for identifying highlights.
Furthermore, the combination of our two audio streams ($A_s + A_d$) yields remarkable performance. This audio-only configuration is nearly on par with the proposed model ($V + A_s + A_d$) and substantially surpasses the conventional bimodal combination of $V + A_s$. These findings strongly indicate that a sophisticated audio representation that integrates both semantic and dynamic features can provide cues for highlight detection.

\vspace{4pt}
\textbf{Audio Fusion Strategy} \; We examine the key design choices within our Audio Feature Fusion module (described in Section~\ref{ssec:audio_fusion}) to verify the effectiveness of our proposed architecture. Our model employs an Early-SA configuration, where self-attention is applied to each auditory stream before fusion, combined with element-wise Multiplication for the fusion operation itself (Eq.~\ref{eq:audio_fusion}). We compare this design against three alternatives: Early-SA with Concatenation, and two Late-SA configurations with Concatenation and Multiplication. In the Late-SA configuration, self-attention is applied single time, after the initial semantic($\mathbf{Z}_a^s$) and dynamic($\mathbf{Z}_a^d$) features are fused.
As shown in Table~\ref{tab:ablation_fusion}, the results confirm that our proposed configuration (Early-SA with Multiplication) achieves the best performance. The most critical factor is the SA placement, with Early-SA consistently outperforming Late-SA. This finding suggests that modeling temporal context within each audio stream individually is crucial before they are fused. Furthermore, the superiority of Multiplication over Concatenation within the effective Early-SA setting can be attributed to its function as a gating mechanism. A key factor in our model's success is the synergy between pathways, wherein dynamic features effectively modulate semantic features to produce a more discriminative representation. This mechanism allows the model to amplify semantic content that is temporally correlated with salient dynamic events, leading to a more robust highlight detection.

\vspace{-6pt}
\section{Conclusion}
\label{sec:typestyle}
\vspace{-3pt}

In this work, we introduced a novel Dual-Pathway Audio Encoders for Audio-Visual Video Highlight Detection. Our architecture is designed to explicitly model both the high-level semantic content and the low-level spectro-temporal dynamics of the audio signal, addressing a key limitation in prior works where audio is often treated superficially. Our experiments demonstrate the effectiveness of our approach, achieving new state-of-the-art performance on the large-scale Mr.HiSum benchmark. Notably, our audio-only configuration surpassed conventional audio-visual models, underscoring the importance of a sophisticated audio representation.

\noindent
\textbf{Acknowledgements.} This work was partly supported by Institute of Information \& communications Technology Planning \& Evaluation (IITP) grant funded by the Korea government(MSIT)[No. RS-2022-II220320, 20220-00320, 50\%] and [No. RS2022-II220641, 50\%].


\bibliographystyle{IEEEbib}
\bibliography{strings,refs}

@inproceedings{imagenet,
  title={Imagenet: A large-scale hierarchical image database},
  author={Deng, Jia and Dong, Wei and Socher, Richard and Li, Li-Jia and Li, Kai and Fei-Fei, Li},
  booktitle={2009 IEEE conference on computer vision and pattern recognition},
  pages={248--255},
  year={2009},
  organization={Ieee}
}

@article{adam,
  title={Adam: A method for stochastic optimization},
  author={Kingma, Diederik P and Ba, Jimmy},
  journal={arXiv preprint arXiv:1412.6980},
  year={2014}
}

@inproceedings{tvsum,
  title={Tvsum: Summarizing web videos using titles},
  author={Song, Yale and Vallmitjana, Jordi and Stent, Amanda and Jaimes, Alejandro},
  booktitle={Proceedings of the IEEE conference on computer vision and pattern recognition},
  pages={5179--5187},
  year={2015}
}

@inproceedings{resnet,
  title={Deep residual learning for image recognition},
  author={He, Kaiming and Zhang, Xiangyu and Ren, Shaoqing and Sun, Jian},
  booktitle={Proceedings of the IEEE conference on computer vision and pattern recognition},
  pages={770--778},
  year={2016}
}

@inproceedings{inceptionv3,
  title={Rethinking the inception architecture for computer vision},
  author={Szegedy, Christian and Vanhoucke, Vincent and Ioffe, Sergey and Shlens, Jon and Wojna, Zbigniew},
  booktitle={Proceedings of the IEEE conference on computer vision and pattern recognition},
  pages={2818--2826},
  year={2016}
}

@inproceedings{audioset,
  title={Audio set: An ontology and human-labeled dataset for audio events},
  author={Gemmeke, Jort F and Ellis, Daniel PW and Freedman, Dylan and Jansen, Aren and Lawrence, Wade and Moore, R Channing and Plakal, Manoj and Ritter, Marvin},
  booktitle={2017 IEEE international conference on acoustics, speech and signal processing (ICASSP)},
  pages={776--780},
  year={2017},
  organization={IEEE}
}

@article{transformer,
  title={Attention is all you need},
  author={Vaswani, Ashish and Shazeer, Noam and Parmar, Niki and Uszkoreit, Jakob and Jones, Llion and Gomez, Aidan N and Kaiser, {\L}ukasz and Polosukhin, Illia},
  journal={Advances in neural information processing systems},
  volume={30},
  year={2017}
}

@inproceedings{kin400,
  title={Quo vadis, action recognition? a new model and the kinetics dataset},
  author={Carreira, Joao and Zisserman, Andrew},
  booktitle={proceedings of the IEEE Conference on Computer Vision and Pattern Recognition},
  pages={6299--6308},
  year={2017}
}

@inproceedings{3dcnn,
  title={Can spatiotemporal 3d cnns retrace the history of 2d cnns and imagenet?},
  author={Hara, Kensho and Kataoka, Hirokatsu and Satoh, Yutaka},
  booktitle={Proceedings of the IEEE conference on Computer Vision and Pattern Recognition},
  pages={6546--6555},
  year={2018}
}

@inproceedings{tvsum_ref1, 
  title={Learning trailer moments in full-length movies with co-contrastive attention},
  author={Wang, Lezi and Liu, Dong and Puri, Rohit and Metaxas, Dimitris N},
  booktitle={European Conference on Computer Vision},
  pages={300--316},
  year={2020},
  organization={Springer}
}

@article{panns,
  title={Panns: Large-scale pretrained audio neural networks for audio pattern recognition},
  author={Kong, Qiuqiang and Cao, Yin and Iqbal, Turab and Wang, Yuxuan and Wang, Wenwu and Plumbley, Mark D},
  journal={IEEE/ACM Transactions on Audio, Speech, and Language Processing},
  volume={28},
  pages={2880--2894},
  year={2020},
  publisher={IEEE}
}

@article{avslowfast,
  title={Audiovisual slowfast networks for video recognition},
  author={Xiao, Fanyi and Lee, Yong Jae and Grauman, Kristen and Malik, Jitendra and Feichtenhofer, Christoph},
  journal={arXiv preprint arXiv:2001.08740},
  year={2020}
}

@inproceedings{sfauditstream,
  title={Slow-fast auditory streams for audio recognition},
  author={Kazakos, Evangelos and Nagrani, Arsha and Zisserman, Andrew and Damen, Dima},
  booktitle={ICASSP 2021-2021 IEEE International Conference on Acoustics, Speech and Signal Processing (ICASSP)},
  pages={855--859},
  year={2021},
  organization={IEEE}
}

@inproceedings{jointav,
  title={Joint visual and audio learning for video highlight detection},
  author={Badamdorj, Taivanbat and Rochan, Mrigank and Wang, Yang and Cheng, Li},
  booktitle={Proceedings of the IEEE/CVF International Conference on Computer Vision},
  pages={8127--8137},
  year={2021}
}

@article{vatt,
  title={Vatt: Transformers for multimodal self-supervised learning from raw video, audio and text},
  author={Akbari, Hassan and Yuan, Liangzhe and Qian, Rui and Chuang, Wei-Hong and Chang, Shih-Fu and Cui, Yin and Gong, Boqing},
  journal={Advances in neural information processing systems},
  volume={34},
  pages={24206--24221},
  year={2021}
}

@article{mbt,
  title={Attention bottlenecks for multimodal fusion},
  author={Nagrani, Arsha and Yang, Shan and Arnab, Anurag and Jansen, Aren and Schmid, Cordelia and Sun, Chen},
  journal={Advances in neural information processing systems},
  volume={34},
  pages={14200--14213},
  year={2021}
}

@inproceedings{pglsum,
  title={Combining global and local attention with positional encoding for video summarization},
  author={Apostolidis, Evlampios and Balaouras, Georgios and Mezaris, Vasileios and Patras, Ioannis},
  booktitle={2021 IEEE international symposium on multimedia (ISM)},
  pages={226--234},
  year={2021},
  organization={IEEE}
}

@inproceedings{umt,
  title={Umt: Unified multi-modal transformers for joint video moment retrieval and highlight detection},
  author={Liu, Ye and Li, Siyuan and Wu, Yang and Chen, Chang-Wen and Shan, Ying and Qie, Xiaohu},
  booktitle={Proceedings of the IEEE/CVF conference on computer vision and pattern recognition},
  pages={3042--3051},
  year={2022}
}

@inproceedings{eclipse,
  title={Eclipse: Efficient long-range video retrieval using sight and sound},
  author={Lin, Yan-Bo and Lei, Jie and Bansal, Mohit and Bertasius, Gedas},
  booktitle={European Conference on Computer Vision},
  pages={413--430},
  year={2022},
  organization={Springer}
}

@inproceedings{nam2022frequency,
  title={Frequency Dynamic Convolution: Frequency-Adaptive Pattern Recognition for Sound Event Detection},
  author={Nam, Hyeonuk and Kim, Seong-Hu and Ko, Byeong-Yun and Park, Yong-Hwa},
  booktitle={Proc. Interspeech 2022},
  pages={2763--2767},
  year={2022}
}

@article{mrhisum,
  title={Mr. hisum: A large-scale dataset for video highlight detection and summarization},
  author={Sul, Jinhwan and Han, Jihoon and Lee, Joonseok},
  journal={Advances in Neural Information Processing Systems},
  volume={36},
  pages={40542--40555},
  year={2023}
}

@inproceedings{nam2024diversifying,
  title={Diversifying and Expanding Frequency-Adaptive Convolution Kernels for Sound Event Detection},
  author={Nam, Hyeonuk and Kim, Seong-Hu and Min, Deokki and Lee, Junhyeok and Park, Yong-Hwa},
  booktitle={Proc. Interspeech 2024},
  pages={97--101},
  year={2024}
}

@inproceedings{csta,
  title={CSTA: CNN-based spatiotemporal attention for video summarization},
  author={Son, Jaewon and Park, Jaehun and Kim, Kwangsu},
  booktitle={Proceedings of the IEEE/CVF Conference on Computer Vision and Pattern Recognition},
  pages={18847--18856},
  year={2024}
}

@article{nam2025temporal,
  title={Temporal Attention Pooling for Frequency Dynamic Convolution in Sound Event Detection},
  author={Nam, Hyeonuk and Park, Yong-Hwa},
  journal={arXiv preprint arXiv:2504.12670},
  year={2025}
}

@inproceedings{lee2025video,
  title={Video Summarization with Large Language Models},
  author={Lee, Min Jung and Gong, Dayoung and Cho, Minsu},
  booktitle={Proceedings of the Computer Vision and Pattern Recognition Conference},
  pages={18981--18991},
  year={2025}
}

\end{document}